%% file: templateArxiv.tex
\documentclass{article}

\usepackage{PRIMEarxiv}
\usepackage{xspace}
\usepackage[utf8]{inputenc} 
\usepackage[T1]{fontenc}    
\usepackage{hyperref}       
\usepackage{url}            
\usepackage{booktabs}       
\usepackage{amsfonts}       
\usepackage{nicefrac}       
\usepackage{microtype}      
\usepackage{lipsum}
\usepackage{fancyhdr}       
\usepackage{graphicx}       
\usepackage{comment}
\graphicspath{{media/}}     

\title{\textsc{Readle}: A Formal Framework for Designing AI-based Edge Systems}

\author{
  Aftab Hussain \\
  University of Houston \\
  Houston, Texas, USA\\
  \texttt{ahussain27@uh.edu} \\
}

\begin{document}
\maketitle
\input{texstuff.tex}

\begin{abstract}

With the wide spread use of AI-driven systems in the edge (a.k.a edge intelligence systems), such as autonomous driving vehicles, wearable biotech devices, intelligent manufacturing, etc., such systems are becoming very critical for our day-to-day lives. A challenge in designing edge intelligence systems is that we have to deal with a large number of constraints in two design spaces that form the basis of such systems: the \textit{edge design space} and the \textit{deep learning design space}. Thus in this work, a new systematic, extendable, manual approach,~\rdl, is proposed for creating representations of specifications in edge intelligent systems, capturing constraints in the edge system design space (e.g. timing constraints and other performance constraints) and constraints in the deep learning space (e.g. model training duration, required level of accuracy) in a coherent fashion. In particular,~\rdl leverages benefits of real-time logic and binary decision diagrams to generate unified specifications. Several insights learned in building \rdl are also discussed, which should help in future research in the domain of formal specifications for edge intelligent systems.
\end{abstract}


\input{sections/intro}
\input{sections/back}

\input{sections/method}

\input{sections/discussion}
\input{sections/related}

\input{sections/conclusion}

\section*{Acknowledgments}
This work is part of a project done for the Fall 2022 Real-time Systems Course at the University of Houston under the supervision of Professor Albert Cheng, who provided useful feedback for this work and also provided beneficial guidance towards understanding the core foundations of this work, and the motivations behind them.

\bibliographystyle{unsrt}
\bibliography{sample-base}

\end{document}

%% file: texstuff.tex
\newcommand{\Fix}[1]{\textbf{\textcolor{red}{Fix/TODO}: #1}}
\newcommand{\Ans}[1]{\textbf{\textcolor{blue}{Answer}: #1}}
\newcommand{\Part}[1]{\noindent\textbf{#1}}
\newcommand{\fsize}[2]{{\fontsize{#1}{0}\selectfont#2}}
\newcommand{\crossmark}{$\times$}
\newcommand{\Space}[1]{}

\newcommand{\eg}{\textit{e.g.}\xspace}
\newcommand{\ie}{\textit{i.e.}\xspace}
\newcommand{\Fone}{\textsc{$F_1$-Score}\xspace}

\newcounter{observation}
\newcommand{\observation}[1]{\refstepcounter{observation}
        \begin{center}
        \vspace{2pt}
        \Ovalbox{
        \begin{minipage}{0.9\columnwidth}
            \textbf{Observation \arabic{observation}:} #1
        \end{minipage}
        }
        \end{center}
}

\newcommand{\MT}{\ensuremath{t}}
\newcommand{\OT}{\ensuremath{t_o}}
\newcommand{\RT}{\ensuremath{t_r}}

\newcommand{\vm}{\textsc{VarMisuse}\xspace}
\newcommand{\mnp}{\textsc{MethodName}\xspace}

\newcommand{\JS}{\textsc{Java-Small}\xspace}
\newcommand{\JM}{\textsc{Java-Med}\xspace}
\newcommand{\JL}{\textsc{Java-Large}\xspace}
\newcommand{\JLR}{\textsc{Java-Large-Roles}\xspace}
\newcommand{\JLT}{\textsc{Java-Large-Transformed}\xspace}
\newcommand{\JLTR}{\textsc{Java-Large-Transformed-Roles}\xspace}
\newcommand{\PY}{\textsc{Py150-Great}\xspace}
\newcommand{\SA}{\textsc{Sorting-Algorithm}\xspace}
\newcommand{\JTT}{\textsc{Java-Top10}\xspace}

\newcommand{\nmc}{{neural model of code}\xspace}
\newcommand{\nmcs}{{neural models of code}\xspace}
\newcommand{\ctv}{\textsc{Code2Vec}\xspace}
\newcommand{\cts}{\textsc{Code2Seq}\xspace}
\newcommand{\rnn}{\textsc{RNN}\xspace}
\newcommand{\tra}{Transformer\xspace}
\newcommand{\ggnn}{\textsc{GGNN}\xspace}
\newcommand{\great}{\textsc{Great}\xspace}
\newcommand{\rdl}{\textsc{Readle}\xspace}

\newcounter{magicrownumbers}
\newcommand\rnum{\stepcounter{magicrownumbers}\arabic{magicrownumbers}}

\newcommand{\basicalert}[2]{\fbox{\bfseries\sffamily\scriptsize\color{blue} #1}{\sf\small$\blacktriangleright$\textit{\color{red} #2}$\blacktriangleleft$}}
\newcommand{\bowen}[1]{\basicalert{From Bowen}{#1}}

%% file: sections/intro.tex
\section{Introduction}
\label{sec-intro}

With the rapid advances in AI towards making fast and accurate predictions, we are today witnessing the adoption of AI in a wide range of application domains, such as autonomous driving systems, unmanned aerial vehicles (UAVs), voice assistants in mobile phones, and even software code development~\cite{alon2018code2seq,alon2018code2vec,aiedge,uav-dl}. The key reason behind the growth of AI is the increasingly growing capabilities of deep learning (DL) systems, also referred to as deep neural networks (DNNs). Today, advances in storage size and computation power of computing systems has made it possible to train very large DNNs (containing in the order of billions of neurons), which could thereby be used to make accurate inferences (or predictions) in a wide range of domains. 

An important application domain where deep learning is rapidly being adopted is edge systems~\cite{trend}. Edge systems comprise of computing systems being put close to data generators, e.g. sensors, in order to enable rapid processing of data locally. In contrast, client-server based systems store the majority of the data in servers, which require transmission of data between the client and servers -- hence creating a delay in getting an output from the computation of an input data. While these delays may still be in the order of seconds, in many mission-critical scenarios, for example autonomous driving or automated industrial assembly, such delays may be costly leading to catastrophic consequences. Edge systems thereby reduce this transmission overhead by providing a temporary data storage (via caches) and a computation platform close to the data producers~\cite{aiedge}. Consequently, edge systems have been adopted in a wide spectrum of application areas including, virtual reality, 3D printers, self-driving cars, robots, intelligent manufacturing, etc.

The quest for further improving edge systems have led practitioners to bring AI to the edge, leading to the development of \textit{edge intelligence systems}. However, building such systems has been challenging. Deep learning algorithms used in DNNs are highly compute-intensive, whereas edge systems are constrained by compute resources. Furthermore, depending on the application context, edge systems have specific functional constraints, e.g. timing constraints and security requirements. (We look in further detail into the operation of DNNs and the constraints of edge systems in Sections~\ref{subsec-back-dl} and~\ref{subsec-back-edge}, respectively). 

While many approaches have been proposed to bridge the computation gap between DNNs and edge systems, some of which we explore in Section~\ref{sec-related}, choosing the right approach for a given edge system can be a challenging task. As we explain in Section~\ref{method-rtl-dnn}, DNNs can have a wide range of configurations and settings under which they are trained, and thus finding the right fit for a given edge system is difficult. Moreover, different edge systems come with their own share of multifarious constraints and requirements, adding further complexity to the problem of finding the right DNN for a given edge system.

We thus focus on the following \textbf{research problem}: \textit{How can we clearly reason about our design decisions in developing an edge intelligence system?} The complexity of choices and restrictions in the two design paradigms of an edge intelligence system (the edge design space and the DNN design space) calls for a need to systematically map the edge intelligence system design process in some form of a common medium. In addition, it has often been seen that when disciplines operate independently of each other towards designing any system, major deficiencies may arise~\cite{donnorman}.  

To address this, we thus develop a novel formal systematic framework, \rdl, geared towards representing specifications of edge intelligence systems. \rdl is based on Real-time Logic (RTL)~\cite{rtl} and Binary Decision Diagrams (BDDs)~\cite{bdd1959,bdd1978}, reaping benefits from both representation systems. RTL is useful for creating specifications of event-action models (models that capture data dependency and temporal ordering of computational actions that should be taken when certain events occur within a real-time system~\cite{rtl-book}). The main benefit of RTL representations is that they can be mechanically manipulated, and thus are amenable for automated processing and verification~\cite{rtl-book}. Furthermore, they are also easily readable and widely accepted by system design practitioners. On the other hand, BDDs offer compactness -- they are concise graphical representations of Boolean logic formulas~\cite{rtl-book}. While formal methods have been previously used for verifying DL systems, e.g.~\cite{bdd,formal}, in this work, we focus on how to build a clear design pathway for building DL systems in the edge (edge intelligece systems.) 

The main motivation behind proposing \rdl was to provide a coherent, unified representation of the design specifications in designing edge intelligence systems. Given the different characteristics of the two underlying design spaces of edge intelligence systems, where the edge space is more event-driven than the DNN design space, building \rdl required appropriately mapping features of DNNs into an event-based context. This allowed the use of RTL to represent constraints in both the design spaces, and thus create unified representations of the entire edge intelligence system.  

\noindent \textbf{Contributions.} Overall we make the following contributions in this paper:
\begin{itemize}
    \item We present a new systematic, extendable framework,~\rdl, for creating for representations of specifications in edge intelligence systems.
    \item \rdl represents constraints in the edge system design space (e.g. timing constraints and other performance constraints) and constraints in the deep neural network space (e.g. training duration, required level of accuracy) in a coherent fashion.
    \item We present several insights in building \rdl that should help in future development of formal representations of specifications in the space of edge intelligence systems and even other AI-equipped systems bounded by real-time systems and DNN constraints. 
\end{itemize}

\noindent \textbf{Paper Organization.} The rest this paper is organized as follows: In Section~\ref{sec-back}, we discuss the preliminary ideas behind \rdl, in particular RTL, BDDs, edge intelligence systems, and deep learning systems. In Section~\ref{sec-method}, we describe the \rdl approach, and subsequently, in Section~\ref{sec-disc}, provide a discussion about \rdl and possible future directions. In Section~\ref{sec-related}, we outline some works in DNN optimization for edge systems, where \rdl may be applied, and also present works that have used formal methods for verifying DL systems. We conclude the paper in Section~\ref{sec-conclusion}.


%% file: sections/back.tex
\section{Preliminaries}
\label{sec-back}

In this section, we describe the notations of the two building blocks of~\rdl: RTL (Subsection~\ref{subsec-back-rtl}) and BDDs (Subsection~\ref{subsec-back-bdd}). In addition, we present some relevant background knowledge on edge systems (Subsection~\ref{subsec-back-edge}) and deep neural networks (Subsection~\ref{subsec-back-dl}). 

\subsection{Real-Time Logic (RTL)}
\label{subsec-back-rtl}

RTL follows the notations of first order logic, with additional features that capture event based timing constraints of an event-action model~\cite{rtl-book}. This makes them very appropriate for modelling most edge systems, for example autonomous driving systems. The notations are described below, following the definitions in~\cite{rtl-book}:

\begin{itemize}
    \item $@(...)$ - an occurrence function that assigns time values to event occurrences. For example, $@(ForkApproach,i) = x$, means the $i$th instance of the event $ForkApproach$ (a driver-less car approaches a fork while commuting on a freeway) in an autonomous driving system, occurs at time $x$.
    \item $@(\uparrow...)$ - If an $\uparrow$ precedes the event name, the event represents the beginning of an action, e.g. $@(\uparrow ReduceSpeed)$
    \item $@(\downarrow...)$ - If a $\downarrow$ precedes the event name, the event represents the end of an action. 
\end{itemize}

\subsection{Binary Decision Diagrams (BDDs)}
\label{subsec-back-bdd}

BDDs are an improved way of representing Boolean logic formulas, over former representations like truth tables, Karnaugh maps, and canonical sum-of-products, which can contain an exponential number of redundant entries~\cite{rtl-book}. BDDs remove the redundancy by representing a Boolean logic formula in a binary decision tree, where: 
\begin{itemize}
    \item the \textit{non-leaf nodes} correspond to variables in the Boolean logic formula,  
    \item each non-leaf node has two \textit{outgoing edges} labeled with 0 or 1 (which correspond to the truth value of the variable), and
    \item there are two \textit{leaf nodes} labelled with 0 or 1, that correspond to the truth value of the entire formula.
\end{itemize}

\begin{figure}
\centering
  \includegraphics[width=45mm]{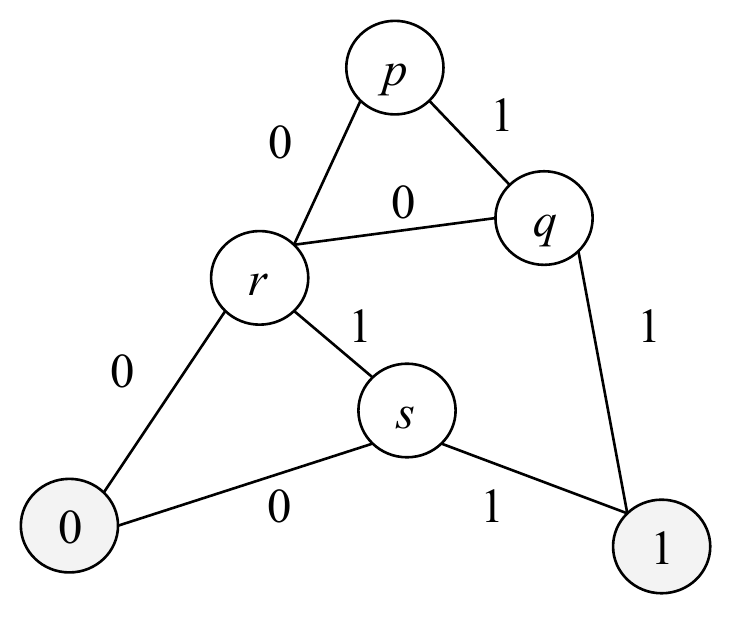}
  \caption{BDD of the formula: $(p \wedge q) \vee (r \wedge s)$.}
  \label{fig:bdd}
\end{figure}

Figure~\ref{fig:bdd}, shows an example of a BDD for a simple Boolean formula. Starting from the root variable node, we can incrementally deduce the value of the formula by following either of the two outgoing edges (that correspond to a 0 or 1 input value) of each variable node that we encounter, until we reach a leaf node (shown in gray in Figure~\ref{fig:bdd}).

\subsection{Edge Systems}
\label{subsec-back-edge}

The edge (or an edge system) refers to a distributed computing paradigm that brings data computation and storage closer to the sources of data, which obviates the need of transferring data to perform computations on them, and thus improves response times~\cite{edge-computing}. An insightful example of an edge system is an autonomous vehicle, which consists of hundreds of sensors collecting data for collision detection. In such systems, time is of the essence in making decisions, and thus all data processing happens locally within the autonomous driving system, rather in a cloud server (which would entail costly delays). 

Edge systems can have a wide range of performance requirements such as low latency and security. Some performance requirements arise from limitations in available resources of edge systems, for example the need to be energy efficient due to limited availability of battery power. We show how these requirements can be captured in Section~\ref{sec-method}.


\subsection{Deep Neural Networks (DNNs)}
\label{subsec-back-dl}

In a nutshell, a DNN model takes in set of input values, e.g. pixel values of an image, and makes an \textit{inference or prediction}, for e.g. whether an image consists of a given stop sign or not. Before a DNN can make inferences on unseen data, it goes through a \textit{training phase}, where it learns from existing datasets about patterns in the dataset. The training phase yields a pretrained version of the model. Typically, this phase is carried out in a local support system, outside the edge system -- only the trained model is loaded into the edge system to perform inferences. This is because training can be costly, where a model is trained with very a large dataset over several iterations (or training steps), each of which can take up to several hours. 

In order to understand the different ways in which a DNN can be configured (as elaborated upon in Section~\ref{method-rtl-dnn}), let us take a look at some of the internal components of a DNN.

\begin{figure}
\centering
  \includegraphics[width=45mm]{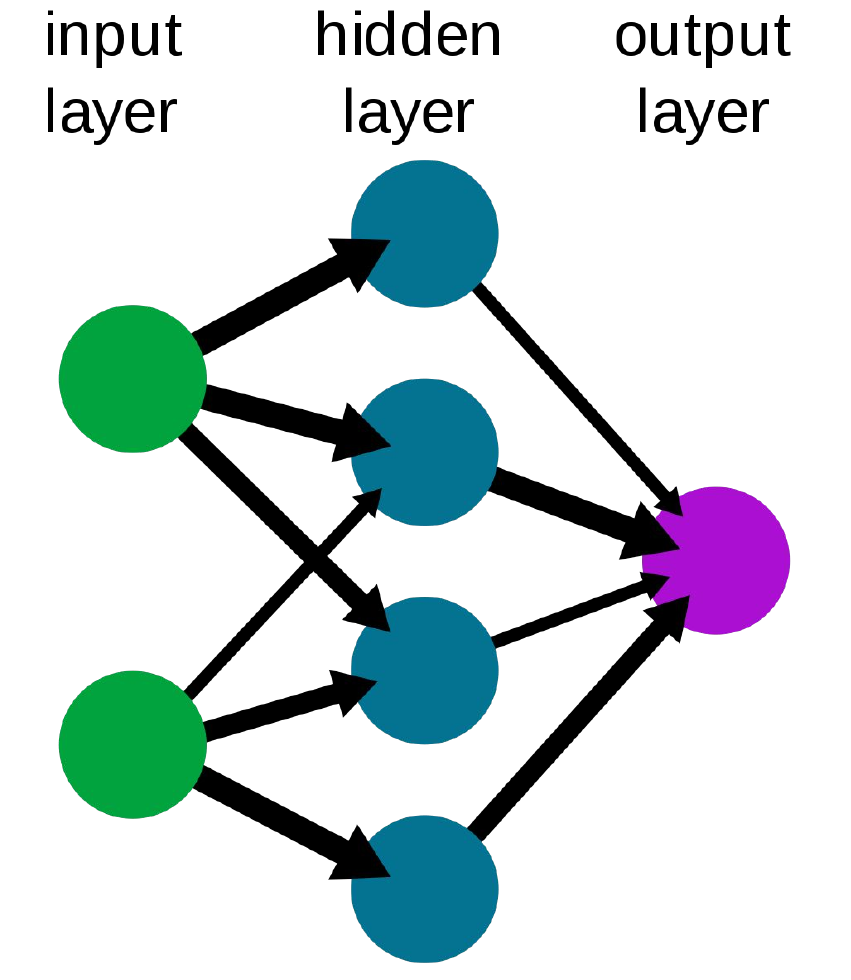}
  \caption{A simple neural network~\cite{simplenn}.}
  \label{fig-simplenn}
\end{figure}

A DNN consists of layers of neurons (mathematical functions), interconnected with each other via weighted edges. The values in these weighted edges are learned during the training phase. Each neuron performs specific operations on the input values it receives. Figure~\ref{fig-simplenn} shows a simple neural network, consisting of an input layer, a hidden layer, and an output layer. DNNs extend the design of a simple neural network by adding more layers, and more neurons in each layer (in the order of billions), and hence can get computationally very expensive (by requiring a high order of floating point operations). The manner in which the layers of neurons are connected with each other and perform these computations is what distinguishes one DNN from the other. For example, computer vision AI applications, which have been widely used in edge systems~\cite{aiedge}, use Convolutional Neural Networks (CNNs). CNNs apply a mathematical operation called convolution in some of its layers~\cite{deeplearningbook}.

%% file: sections/method.tex
\section{The \rdl Approach}
\label{sec-method}

In this section, we first present the workflow of the proposed approach, \rdl, Then we identify features of the edge design space and the DNN design space, explaining how the design features can be captured using \rdl (Subsections~\ref{method-rtl-edge}, \ref{method-rtl-dnn}).

\subsection{Overall Methodology}
\label{method-ov}

\begin{figure*}
\centering
  \includegraphics[width=140mm]{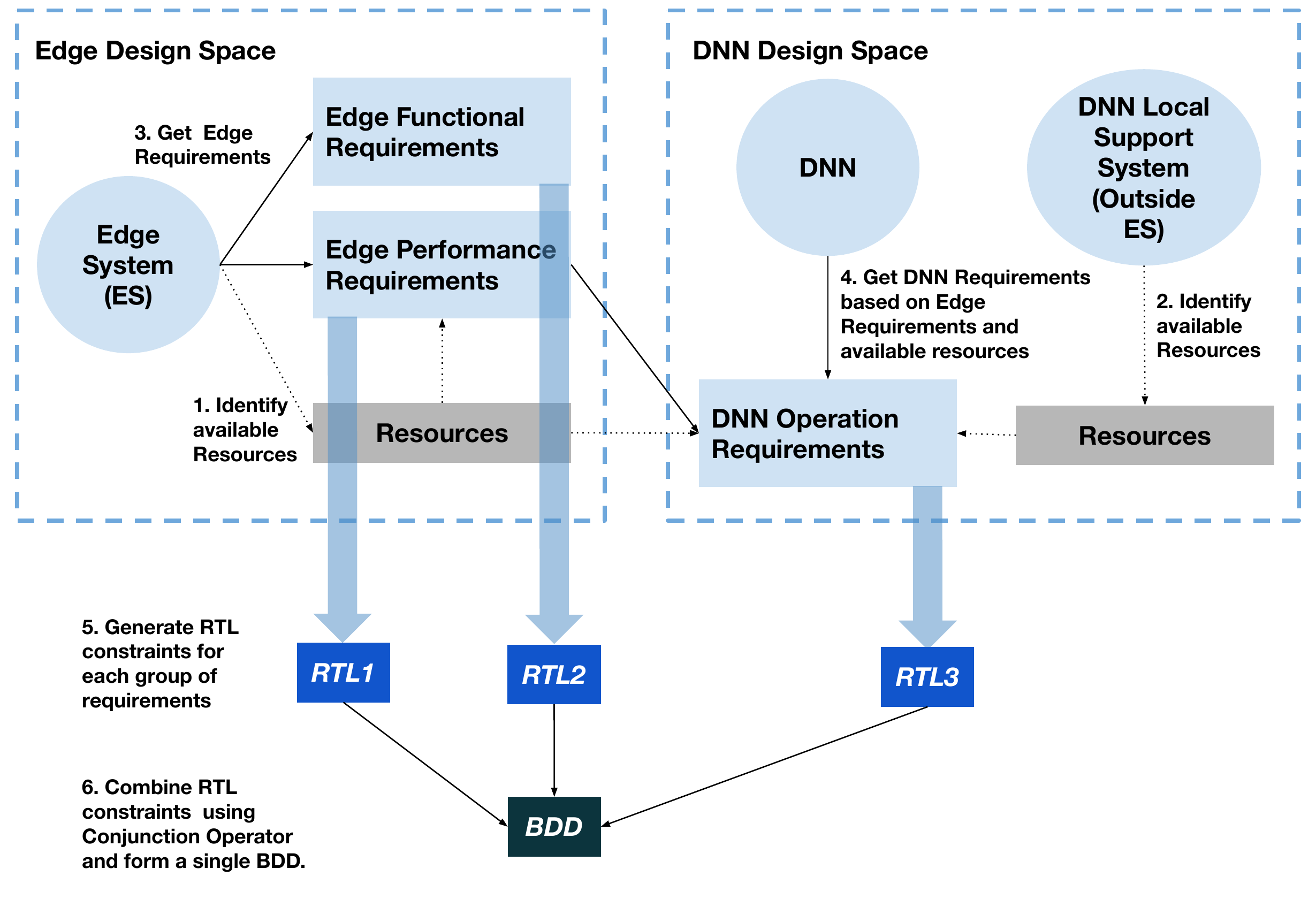}
  \caption{The \rdl Methodology}
  \label{fig-readle}
\end{figure*}

Figure~\ref{fig-readle}, shows the overall workflow of \rdl. The first step involves identifying available resources in the two design spaces, for example, storage memory, battery lifetime, etc. Next, depending on these resources we specify the requirements, in English, for each design space. Since the DNN model resides in the edge system, some of the requirements and resources in the edge space also affects the requirements in the DNN design space. Once all requirements are specified in both spaces, RTL specifications are generated, and then finally combined by conjunction (AND operator), from which a BDD is constructed in the manner explained in Section~\ref{subsec-back-bdd}. The key attribute of \rdl lies in the fact that allows systematically connecting related requirements in two design spaces, and obtain a unified representation of the specification deduced from two design spaces. We next go into further details into how the requirements in the two spaces are captured, in the subsequent subsections.

\subsection{Capturing Requirements in the Edge Design Space}
\label{method-rtl-edge}

As we can see from Figure~\ref{fig-readle}, there are two types of requirements in the edge design space: functional and performance requirements. 

\begin{itemize}
    \item \textit{Functional requirements (FRs)} specify what the system should do given an event, e.g., if an autonomous vehicle comes close to a red traffic light, it should stop -- In \rdl, this specification can be stated as follows $\forall x@(RedLightApproach,x) <= @(\uparrow BraketoStop, x)$.
    \item \textit{Performance requirements (PRs)} are those that follow temporal or energy consumption constraints, e.g. how fast should an autonomous vehicle react when witnessing a stop light, or how an autonomous car can control its acceleration without consuming power greater than a given threshold? A sample performance requirement is that a car battery needs at least $P$ Watts of power to come to a safe stop -- In \rdl, this specification can be stated as $forall x @(PowerBelow: P + threshold, x) <= @(\uparrow TriggerUnsafetoDriveWarning, x)$. Notice how in \rdl we modified the use of RTL to incorporate non-temporal constraints. 
\end{itemize}

As shown in Figure~\ref{fig-readle}, the resources available governs the performance requirements of a system. Consequently, performance requirements (e.g. required accuracy of predictions, maximum allowable down times) and resources also affects the operation requirements of the pretrained DNN model, which resides in the edge system, which we observe next.

 \subsection{Capturing Requirements in the DNN Design Space}
\label{method-rtl-dnn}

Requirements in the DNN design space arise from a number of decisions related to training settings and model parameters that need to be considered while deploying DNNs in a given edge system, such as an autonomous vehicle. These decisions are governed primarily by two application-context dependent factors -- (1) the \textit{availability of resources} (e.g. storage memory, processor capacity, battery power), and the (2) the \textit{performance requirements} from the edge space. We present some of those training settings and parameters below that can be specified by the \rdl approach:

\begin{itemize}
    \item \textit{Training duration.} The length of training (number of training steps) determines to a large extent the quality of the predictions of a model, the longer the training, the better the performance. As previously mentioned, training is supported by a system outside the edge system, due to the long duration training can take, and is typically performed when the edge system is down. Thus while setting the training duration, it is less than the maximum allowable downtime of the edge system. In \rdl, we can state this requirement as follows: $forall x @(\downarrow TrainModel,x) - (\uparrow TrainModel,x) <= forall x @(\downarrow SystemDown,x) - @(\uparrow SystemDown,x) + ModelLoadTime$. Edge systems can also stipulate a certain level of accuracy that a trained model must meet in making predictions. 
    \item \textit{Training batch size.} Bigger training batch sizes, allow for more training data to be processed through a model during training during each iteration, and hence increases the training speed. However, this batch size is restricted by the amount of memory size available in the system that hosts the training process.
    \item \textit{Size of the Model.} This parameter refers to the total number of neurons in the model, which can be controlled by specifying the number of nodes in each layer of the model. The amount of available storage in the edge is the main constraint that governs how big a pretrained model can be. 
\end{itemize}

%% file: sections/discussion.tex
\section{Discussion and Future Directions}
\label{sec-disc}


\textit{Composability of \rdl.} The first step in the \rdl approach of identifying resources can be wide ranging, given the vast number of functions an edge system (for example an autonomous driving system) can perform. To tackle this scenario the edge system design can be divided into separate spaces based on the function it performs, and then \rdl can be applied to each subdivision. Since \rdl allows compositional construction of constraints, specifications for all the subdivisions can be combined under a single BDD using \rdl.

 \textit{Codifying noise and attacks on DNNs into specifications.} Deep neural networks can also exhibit interesting behaviour when faced with noisy training data (in which case it can memorize the data~\cite{rabin2021memorization, arpit}), or even malicious behaviour when faced with adversarial examples. For example, it has been found that a DNN can mis-classify stop-signs when a yellow Post-it note attached to them as speed-limit signs~\cite{badnet}. Such scenarios can clearly affect the outcome of DNNs in edge systems, and thus we need to explore how checks against such scenarios can be represented in the design specifications of edge intelligence systems.
 
 \textit{User studies to assess cognitive burden.} For a more comprehensive analysis of how well \rdl achieves its goal of providing coherent, unified representations of edge intelligence system design specifications in designing, there is a need for systematic user studies of applying \rdl on designing edge intelligence systems in the real-world, with a focus on measuring the cognitive burden of the designers.
 

%% file: sections/related.tex
\section{Related Works}
\label{sec-related}


\textbf{DNN optimizations for the edge that \rdl can potentially capture.} Arising from the limited resources of edge systems, and the huge computation costs of DL systems, there have been several optimizations made to using deep learning systems, so that they may be feasibly deployed in the edge. These optimizations entail modifications in the DNN which may also be codified in the \rdl framework. Examples of such optimizations include network pruning approaches e.g.~\cite{brain,prune}, which remove irrelevant weights in DNN models that have too many redundant neurons. Other techniques have deployed adaptive NNs, e.g.~\cite{adaptive1,adaptive2}, which skip some layers of the NN to perform computations, for easy inputs. For example BranchyNet~\cite{adaptive2} uses early-exit to skip layers of the NN for easy input images. Quantization techniques address the large memory requirements of large DNNs by encoding the model weights with lower precision, without significantly compromising the accuracy of the model, (e.g. by converting weights and activations from a 32-bit floating point representation level to a 16-bit floating point representation level)~\cite{quant-mobile,trained-uniform}. For interested readers, an elaborate survey on DNN optimization techniques for the edge is provided in~\cite{aiedge,empower-edge}.

\textbf{Use of Formal methods for verification of DL systems}. Gehr et al.~\cite{ai2} present AI2, a scalable analyzer based on abstract interpretation framework~\cite{aif} for deep neural networks, which can automatically prove safety properties (e.g., robustness) of DNNs like convolutional neural networks (CNNs). They define abstract transformers that capture the behavior of common neural network layers (e.g. max pooling and convolutional layers in a CNN). Scheibler et al.~\cite{scheiber} use bounded model checking technique for verifying NNs. Huang et al.~\cite{marta} showed a SMT solver based verification framework for feed forward multilayer NNs. Guy et al.~\cite{reluplex} extend the simplex algorithm for solving linear programming instances in order to verify ReLU (Rectified Linear Unit) activation function constraints in NNs.


%% file: sections/conclusion.tex
\section{Conclusion}
\label{sec-conclusion}

This work presents~\rdl, a new approach for creating representations of specifications in edge intelligence systems, capturing constraints in the edge system design space (e.g. timing constraints and other performance constraints) and constraints in the deep learning space (e.g. model training duration, required level of accuracy) in a coherent fashion. \rdl leverages benefits of real-time logic and binary decision diagrams to generate unified specifications. \rdl is extendable, and may be used to represent multifarious settings (at different levels of abstraction) of building and using DNNs in edge systems. This work thus serves as an idea/perspective paper, which should help in future research towards the development of coherent design approaches for building edge intelligence systems.